\documentstyle[12pt]{article}
\newcommand{\be}{\begin{equation}}
\newcommand{\ee}{\end{equation}}
\begin{document}

\begin{center}
{\bf Chiral Solitons in Generalized Korteweg-de Vries
Equations\footnote{This work is supported in part by funds provided
by the U. S. Department of Energy (D.O.E.) under cooperative research
agreement DE-FC02-94ER40818, and by Conselho Nacional de Desenvolvimento
Cient\'\i fico e Tecnol\'ogico, CNPq, Brazil.}}
\end{center}

\begin{center}
D. Bazeia$^{a,b}$ and F. Moraes$^c$
\end{center}

\begin{center}
$^a$Center for Theoretical Physics\\
Laboratory for Nuclear Science and Department of Physics\\
Massachusetts Institute of Technology, Cambridge, Massachusetts 02139-4307
\end{center}

\begin{center}
$^b$Departamento de F\'\i sica, Universidade Federal da Para\'\i ba\\
Caixa Postal 5008, 58051-970 Jo\~ao Pessoa, Para\'\i ba, Brazil
\end{center}

\begin{center}
$^c$Departamento de F\'\i sica, Universidade Federal de Pernambuco\\
50670-901 Recife, Pernambuco, Brazil
\end{center}

\begin{center}
MIT-CTP-2713
\end{center}

\begin{center}
Abstract
\end{center}

Generalizations of the Korteweg-de Vries equation are considered,
and some explicit solutions are presented. There are situations where
solutions engender the interesting property of being chiral, that is,
of having velocity determined in terms of the parameters that define the
generalized equation, with a definite sign.

\newpage

Recent works \cite{ben96,jac96} have shown that the nonlinear Schr\"odinger
equation presents chiral solitons when nonlinearity enters the game via
derivative coupling. The nonlinear Schr\"odinger equation that appears in
the above works is obtained from a very interesting dimensional reduction to
one space dimension of a planar model \cite{jpi90} describing non-relativistic
matter coupled to a Chern-Simons gauge field. This equation can be cast to the
form
\begin{equation}
iu_t+\lambda\,j\,u+\mu\,u_{xx}=\frac{dV}{d\rho}\,u~,
\end{equation}
where $j=-i\nu(u^{*}u_x -u\,u^{*}_x)$ is the current density, $\lambda,\mu,\nu$
are real parameters and $V=V(u^*u)=V(\rho)$ is the potential, expressed in
terms of the charge density. This equation should be contrasted with
\begin{equation}
iu_t+\mu\,u_{xx}=\frac{dV}{d\rho}\,u~,
\end{equation}
which is the standard nonlinear Schr\"odinger equation -- recall that
$V(\rho)$ is usually quadratic or cubic in $\rho$.

For travelling waves in the form $u(x,t)=\rho(x-ct)\exp[i\theta(x,t)]$ one
finds solutions to the above nonlinear derivative Schr\"odinger equation that
present velocity restricted to just one sense, and the system is then chiral.
This is nicely illustrated in a more recent work on the same subject
\cite{gse97}, where the soliton structure for vanishing and non-vanishing
boundary conditions are investigated. Evidently, the chiral solitons found in
these works may play important role within the context of the fractional
quantum Hall effect, where chiral excitations are known to appear.

As one knows, however, the Korteweg-de Vries \cite{kdv} or KdV equation
is considered the standard equation for soliton solutions, and the mKdV
equation modifies the KdV equation by considering next higher nonlinearity
\cite{whi74,das89,deb97}. They are usually considered for describing
propagation of waves in several distinct physical situations such as for
instance water waves, ion-acoustic waves in plasmas and acoustic waves in
anharmonic crystals. Within this context, it seems interesting to extend the
presence of chiral solitons to this new scenario where the KdV equation plays
the basic role. This is the main subject of the present work, and here we
shall consider specific generalizations of the KdV equation that present
travelling waves that engender the same chiral behavior recently found
in nonlinear derivative Schr\"odiger equations \cite{ben96,jac96,gse97}.

To offer the possibility of extending the presence of chiral solitons to
the scenario where the KdV equation plays the central role, let us start
by recalling that the KdV equation combines dispersion and nonlinearity to
describe solitons in distinct branches of nonlinear science. The generalized
KdV equation that we consider can be written as
\begin{equation}
\label{eq:gKdV}
u_t + f_x- \delta\,u_{xxx}=0~,
\end{equation}
where $f=f(u)$ is a smooth function of $u$. The simplest case is
$f(u)=(1/2)a\,u^2$, which reproduces the original KdV equation. Other
cases are known, one is given by $f(u)=4\,a\,u^3$, which gives
\begin{equation}
\label{eq:phi4}
u_t + 12\,a\,u^2\,u_x - \delta\,u_{xxx}=0~,
\end{equation}
and reproduces the modified KdV or mKdV equation, and another one is given
by $f(u)=4\,a\,u^3-a'\,u^5$, which gives
\begin{equation}
\label{eq:phi6}
u_t+ (12\,a\,u^2-5\,a'\,u^4)\,u_x-\delta\,u_{xxx}=0~.
\end{equation}
Former investigations \cite{dey86} of the above generalized KdV equation
have already been done, but here we shall present novelties when we restrict
attention to odd functions $f(u)=-f(-u)$. The novelties are directly
related to ideas recently introduced in \cite{ben96,jac96},
where chiral solitons have been found. Like the KdV equation,
the above generalized KdV equation also combines dispersion,
directly related to $\delta$, and nonlinearity, which is now
controlled by the nonlinear function $f(u)$. Because we are considering odd
nonlinear functions that go beyond the case governed by the mKdV equation,
we see that one is relaxing the assumption of weak nonlinearity contained in
the standard case.

In general, solutions to the generalized KdV Eq.~{$(\ref{eq:gKdV})$} in
the form of travelling wave $u(x,t)=u(x-ct)=u(y)$ obey
\begin{equation}
\label{eq:field}
\frac{d^2 u}{dy^2}=-\frac{c}{\delta}\,u+\frac{1}{\delta}\,f(u)~,
\end{equation}
if one requires that the resulting equation does not break the discrete
symmetry $u\to-u$. This result is interesting {\it because} we can use it
to map solutions to the generalized KdV equation to solutions of equations
of motion describing relativistic $1+1$ dimensional systems governed
by the Lagrangian density
\begin{equation}
\label{eq:lag}
{\cal L}=\frac{1}{2}\frac{\partial\phi}{\partial x^{\alpha}}
\frac{\partial\phi}{\partial x_{\alpha}}+\frac{v}{2\delta}\phi^2-
\frac{1}{\delta}\int^{\phi}f(\phi')d\phi'~,
\end{equation}
where $x^{\alpha}=(x^0=t,x^1=x)$ and $x_{\alpha}=(x_0=t,x_1=-x)$ -- see
\cite{raj82} for further details. Different systems can be introduced,
and some known examples that present interesting topological solutions are
the $\phi^4$, the $\phi^6$, and the sine-Gordon systems. For instance, the
$\phi^4$ model is directly related to the mKdV Eq.~{$(\ref{eq:phi4})$},
the $\phi^6$ model is directly related to the generalized KdV
Eq.~{$(\ref{eq:phi6})$}, and as far as we know the sine-Gordon system is new,
at least within the context of providing solutions to the corresponding
generalized KdV equation.

To present explicit investigations, let us consider the $\phi^4$ model
as a first example. In this case we have to deal with the mKdV
Eq.~{$(\ref{eq:phi4})$}, and for $u(x,t)=u(x-ct)=u(y)$ we have
\begin{equation}
\frac{d^2 u}{dy^2}=-\frac{c}{\delta}\,u+ 4\frac{a}{\delta}\,u^3~,
\end{equation}
and when $c/\delta>0$, $a/\delta>0$ there are explicit solutions
\begin{equation}
\phi(x,t)=\pm \frac{1}{2}\sqrt{\frac{c}{a}}\,
\tanh\sqrt{\frac{c}{2\delta}\,}\,(x-ct-\bar{x})~,
\end{equation}
where $\bar{x}$ is arbitrary. Note that both the amplitude and width
of the solutions depend on $c$, the velocity, which is bounded
in this case. For these solutions the condition $c/\delta>0$
imposes a definite sign to the velocity and so unveils its chiral behavior.
Let us now consider the $\phi^6$ model as another example. In this case we
have to deal with the generalized KdV Eq.~{$(\ref{eq:phi6})$}, and for
$u(x,t)=u(y)$ we get
\begin{equation}
\frac{d^2 u}{dy^2}=-\frac{c}{\delta}\phi+
4\,\frac{a}{\delta}\,u^3-\frac{a'}{\delta}\,u^5~.
\end{equation}
For $c/\delta<0$, $a/\delta<0$, and $a'/\delta<0$ we have the explicit
solutions
\begin{equation}
u(x,t)=\pm \sqrt{\frac{c}{2a}\,}
\sqrt{1\pm\tanh\sqrt{-\frac{c}{\delta}\,}(x-ct-\bar{x})\,}~.
\end{equation}
These solutions are valid if and only if $c=3a^2/a'$, and so the velocity
is determined by $a$ and $a'$, and carries the same sign $a'$ has, and this
makes the solutions chiral.

We now focus attention on another generalized KdV equation. Here we refer to
the explicit equation
\begin{equation}
u_t+[a+a'\,\cos(b\,u)\,]\,u_x-\delta\,u_{xxx}=0~.
\end{equation}
In this case there are travelling waves $u=u(y)$ that obey
\begin{equation}
\frac{d^2u}{dy^2}= \frac{a'}{b\,\delta}\sin(b\,u)~,
\end{equation}
if we set $c=a$. This is the sine-Gordon equation, which has soliton solutions
\begin{equation}
u(x,t)=\pm\,\frac{2}{b}\,\arctan e^{\sqrt{\frac{a'}{\delta}\,}\,
(x-a\,t-\bar{x})}~.
\end{equation}
We see that the velocity is controlled by $a$ and so the solutions are
chiral. As it is known, the sine-Gordon system is richer than the $\phi^4$
and $\phi^6$ models in providing soliton solutions, and so there are other
solutions as for instance the breather solutions in this case.

The above investigations show that {\it there exist} odd functions $f(u)$
that can be used to change the right hand side of Eq.~{(\ref{eq:field})}
to the form
\begin{equation}
\label{eq:pot1}
-\frac{c}{\delta}\,u+\frac{1}{\delta}\,f(u)=\frac{dV}{du}~,
\end{equation}
where $V=V(u)$ is an even function of $u$. If we now change $u\to\phi$,
that is, if we make the identification $V(u)\to V(\phi)$ and
$dV/du\to dV/d\phi$ we can identify $V(\phi)$ with the potential of the
corresponding relativistic system $(\ref{eq:lag})$. This is the formal map
between travelling solutions to the generalized KdV Eq.~{$(\ref{eq:gKdV})$}
and solutions to relativistic $1+1$ dimensional systems of fields. This form
is interesting since there is a class of relativistic systems defined by
\begin{equation}
\label{eq:pot2}
V(\phi)=\frac{1}{2}\left(\frac{dH}{d\phi}\right)^2~,
\end{equation}
where $H=H(\phi)$ is a smooth function on $\phi$, for which we can further
reduce the degree of the equation of motion. The point here is that the
first-order differential equation
\begin{equation}
\frac{d\phi}{dx}=\frac{dH}{d\phi}~,
\end{equation}
also solve the second-order equation of motion
\begin{equation}
\frac{d^2\phi}{dx^2}=\frac{dH}{d\phi}\,\frac{d^2H}{d\phi^2}~,
\end{equation}
that appears when the potential obeys Eq.~{$(\ref{eq:pot2})$}. Translating to
the generalized KdV equation, we see that the first-order differential equation
\begin{equation}
\frac{du}{dy}=\frac{dh}{du}~,
\end{equation}
solves the generalized KdV Eq.~{$(\ref{eq:field})$} when
\begin{equation}
-\frac{c}{\delta}\,u+\frac{1}{\delta}\,f(u)=
\frac{dh}{du}\frac{d^2h}{du^2}~.
\end{equation}

We recall that the generalized KdV Eq.~{$(\ref{eq:gKdV})$} can be obtained
from a non-local Lagrangian, and symmetry arguments \cite{jac97} show that
Galileo invariance requires the additional boundary condition
$U(x=\infty, t)+U(x=-\infty, t)=0$, where
\begin{equation}
U(x,t)=f(u)-\delta\,u_{xx}~.
\end{equation}
For configurations presenting second space derivative that vanishes
asymptoticaly, and with $u(x=\pm\infty,t)$ giving constant values,
it suffices to impose the condition
\begin{equation}
u(x=\infty,t)+u(x=-\infty,t)=0
\end{equation}
to maintain Galileo invariance -- recall that we are considering odd
$f=f(u)$. As one knows, chiral solitons should violate the above
condition, as happens with the chiral solutions we have introduced above.
Here we take advantage of the connection to relativistic systems to argue
that chiral solutions seem to appear with definite velocity whenever the
generalized KdV equation is related to some relativistic system whose
manifold of vacuum states contains adjacent points that are not connected
by parity symmetry.

The above possibility of writing chiral solutions to generalized KdV equations
can be further generalized to coupled systems of generalized KdV equations.
This is done by first realizing that the term $f_x=(\partial f/\partial x)$
that appears in the generalized KdV Eq.~{$(\ref{eq:gKdV})$} can be used to
provide a natural way of introducing extensions to systems where two or more
configurations interact with each other. In the case of two configurations
$u(x,t)$ and $v(x,t)$, for simplicity, the generalization allows introducing
the pair of equations
\begin{eqnarray}
u_t+ f_x-\delta\,u_{xxx}&=&0~,\\
v_t+ g_x-\bar{\delta}\,v_{xxx}&=&0~,
\end{eqnarray}
where $f=f(u,v)$ and $g=g(u,v)$ now obey $f(u,v)=- f(-u,v)$ and
$f(u,v)=f(u,-v)$, and $g(u,v)=g(-u,v)$ and $g(u,v)=-g(u,-v)$. We remark
that similar systems of coupled modified KdV equations have also been
recently considered in \cite{ihi97} -- see also Ref.~{\cite{his97}} for
investigations concerning the derivative multi-component nonlinear
Schr\"odinger equation.

In the present case we search for travelling solutions $u=u(y)$ and $v=v(y)$
in order to get to
\begin{eqnarray}
\frac{d^2u}{dy^2}&=&-\frac{c}{\delta}\,u+\frac{1}{\delta}\,f(u,v)~,\\
\frac{d^2v}{dy^2}&=&-\frac{c}{\bar{\delta}}\,v+
\frac{1}{\bar{\delta}}\,g(u,v)~,
\end{eqnarray}
as equations that follow from the coupled pair of generalized KdV equations,
obeying the same parity symmetry in $(u,v)$ space as the original equations.
We now introduce $V=V(u,v)$ in order to make the changes
\begin{eqnarray}
-\frac{c}{\delta}\,u+\frac{1}{\delta}\,f(u,v)
&=&\frac{\partial V}{\partial u}~,\\
-\frac{c}{\bar{\delta}}v+\frac{1}{\bar{\delta}}\,g(u,v)
&=&\frac{\partial V}{\partial v}~,
\end{eqnarray}
and here we get to
\begin{eqnarray}
\frac{d^2 u}{dy^2}&=&\frac{\partial V}{\partial u}~,\\
\frac{d^2 v}{dy^2}&=&\frac{\partial V}{\partial v}~.
\end{eqnarray}

Like in the former case this system can also be identified with relativistic
$1+1$ dimensional system, but now describing two coupled real scalar fields.
This kind of systems of two real scalar fields have been already considered
for instance in \cite{raj79} and more recently in \cite{baz95}. Here we
see that the class of sytems introduced in \cite{baz95} is defined
via the specific potential
\begin{equation}
V(\phi,\chi)=\frac{1}{2}\left(\frac{\partial H}{\partial\phi}\right)^2
+\frac{1}{2}\left(\frac{\partial H}{\partial\chi}\right)^2~,
\end{equation}
where $H(\phi,\chi)$ is some smooth function of the two fields $\phi$ and
$\chi$. In this case the equations of motion for static fields are given by
\begin{eqnarray}
\frac{d^2\phi}{dx^2}&=&\frac{\partial H}{\partial\phi}\,
\frac{\partial^2H}{\partial\phi^2}+\frac{\partial H}{\partial\chi}\,
\frac{\partial^2H}{\partial\chi\partial\phi}~,\\
\frac{d^2\chi}{dx^2}&=&\frac{\partial H}{\partial\phi}\,
\frac{\partial^2H}{\partial\phi\partial\chi}+
\frac{\partial H}{\partial\chi}\,\frac{\partial^2H}{\partial\chi^2}~,
\end{eqnarray}
and are solved by the following first-order differential equations
\begin{eqnarray}
\frac{d\phi}{dx}&=&\frac{\partial H}{\partial\phi}~,\\
\frac{d\chi}{dx}&=&\frac{\partial H}{\partial\chi}~.
\end{eqnarray}
This is interesting since we can find travelling wave solutions to specific
pairs of coupled generalized KdV equations by investigating the pair of
first-order equations
\begin{eqnarray}
\frac{du}{dy}&=&\frac{\partial h}{\partial u}~,\\
\frac{dv}{dy}&=&\frac{\partial h}{\partial v}~,
\end{eqnarray}
if it is possible to write
\begin{eqnarray}
-\frac{c}{\delta}\,u+\frac{1}{\delta}\,f(u,v)
&=&\frac{\partial h}{\partial u}\,
\frac{\partial^2h}{\partial u^2}+\frac{\partial h}{\partial v}\,
\frac{\partial^2h}{\partial v\partial u}~,\\
-\frac{c}{\bar{\delta}}v+\frac{1}{\bar{\delta}}\,g(u,v)
&=&\frac{\partial h}{\partial u}\,\frac{\partial^2h}{\partial u\partial v}+
\frac{\partial h}{\partial v}\,
\frac{\partial^2h}{\partial v^2}~,
\end{eqnarray}
for $h=h(u,v)$ smooth in $u$ and $v$.

In \cite{baz95} some systems of two coupled real scalar fields were already
investigated, and explicit solutions were presented. Evidently, we can use
them to present solutions to systems of coupled generalized KdV equations that
present velocity defined by the set of parameters that identify the system,
and this provides other examples where chiral solitons appear. To introduce an
explicit example let us consider the pair of equations
\begin{eqnarray}
u_t+[A-2\,\lambda^2+6\,\lambda^2\,u^2+2\,\mu\,(\lambda+
2\mu)\,v^2]\,u_x\nonumber\\
+4\,\mu\,(\lambda+2\,\mu)\,u\,v\,v_x-u_{xxx}&=&0~,\\
v_t+[A-2\,\lambda\,\mu+6\,\mu^2\,v^2+2\,\mu\,(\lambda+
2\mu)\,u^2]\,v_x\nonumber\\
+4\,\mu\,(\lambda+2\,\mu)\,u\,v\,u_x-v_{xxx}&=&0~,
\end{eqnarray}
where $A$, $\lambda$ and $\mu$ are real parameters. For travelling waves in
the form $u=u(y)$ and $v=v(y)$ we can write, after setting $c=A$,
\begin{eqnarray}
\frac{d^2u}{dy^2}&=&
-2\lambda^2\,u+2\lambda^2\,u^3+2\mu(\lambda+2\mu)\,u\,v^2~,\\
\frac{d^2v}{dy^2}&=&
-2\lambda\mu\,v+2\mu^2\,v^3+2\mu(\lambda+2\mu)\,u^2\,v~.
\end{eqnarray}
These equations are solved by solutions of the first-order differential
equations
\begin{eqnarray}
\frac{du}{dy}&=&\lambda-\lambda\,u^2-\mu\,v^2~,\\
\frac{dv}{dy}&=&-2\,\mu\,u\,v~.
\end{eqnarray}
An explicit pair of solutions is given by, for $\lambda/\mu>2$ for
instance,
\begin{eqnarray}
u(x,t)&=&\tanh[2\mu(x-At-\bar{x})]~,\\
v(x,t)&=&\sqrt{\frac{\lambda}{\mu}-2\,\,}\,\,{\rm sech}[2\mu(x-At-\bar{x})]~,
\end{eqnarray}
and presents chiral behavior.

To summarize we recall that in this paper we have offered a method for
searching for chiral solitons in generalized KdV equations. This method
relies on mapping generalized KdV equations to equations of motion that
appear in relativistic systems of real scalar fields. We have introduced
several examples, and presented explicit travelling solutions that engender
chiral behavior. Evidently, several interesting issues appear very naturally,
one of them is related to the possibility of extending the present
investigations to other nonlinear equations. Investigations on this
\cite{baz98} and in other related issues are now in progress.
\begin{center}
Acknowledgments
\end{center}

We would like to thank Roman Jackiw for interesting comments, and for
reading the manuscript.

\end{document}